\documentclass[a4paper,11pt]{article}
\usepackage{jinstpub} 
\usepackage{lineno}

\title{ML-based Calibration and Control of the GlueX Central Drift Chamber}

\author[a]{T. Britton}
\author[a]{M. Goodrich}
\author[b]{N. Jarvis}
\author[a]{T. Jeske}
\author[a]{N. Kalra}
\author[a,1]{D. Lawrence,\note{Corresponding author.}}
\author[a]{D. McSpadden}
\author[a]{K. Rajput}
\affiliation[a]{Thomas Jefferson National Accelerator Facility, Newport News, VA, USA}
\affiliation[b]{Carnegie Mellon University, Pittsburgh, PA, USA}

\emailAdd{davidl@jlab.org}

\abstract{The GlueX Central Drift Chamber (CDC) in Hall D at Jefferson Lab, used for detecting and tracking charged particles, is calibrated and controlled \textit{during} data taking using a Gaussian process. The system dynamically adjusts the high voltage applied to the anode wires inside the chamber in response to changing environmental and experimental conditions such that the gain is stabilized. Control policies have been established to manage the CDC's behavior. These policies are activated when the model's uncertainty exceeds a configurable threshold or during human-initiated tests during normal production running. We demonstrate the system reduces the time detector experts dedicate to calibration of the data offline, leading to a marked decrease in computing resource usage without compromising detector performance.}


\arxivnumber{2403.13823} 

\begin{document}
\maketitle
\flushbottom

\section{Introduction}

\begin{figure}[htb]
    \centering
    \includegraphics[width=0.55\linewidth]{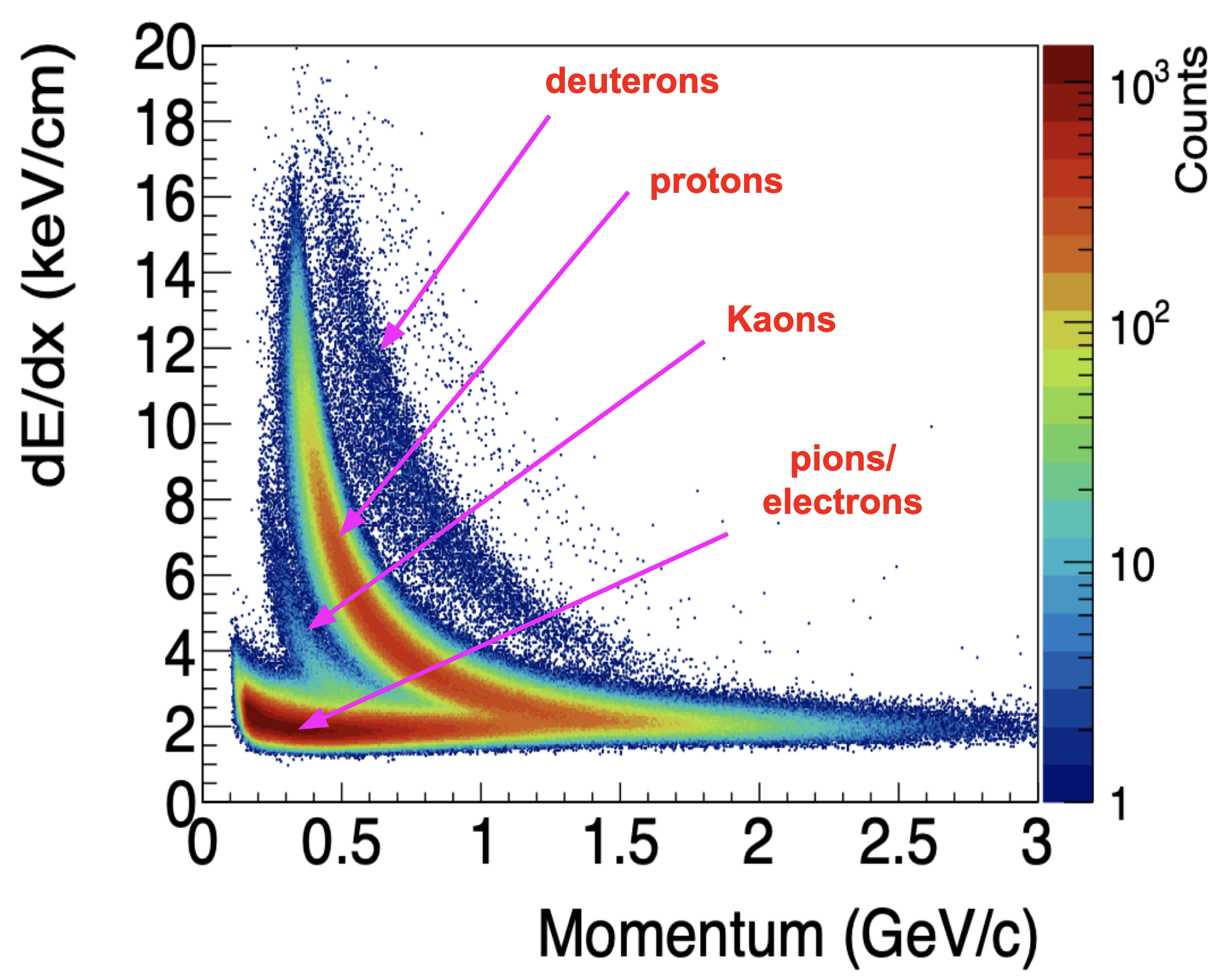}
    \caption{Energy loss rate as a function of momentum in the CDC. For lower momentum particles this can be used for identifying the particle type. Accurate gain calibration helps sharpen the resolution between the various bands.}
    \label{fig:CDC_PID}
\end{figure}

\begin{figure}[htb]
    \centering
    \includegraphics[width=0.9\linewidth]{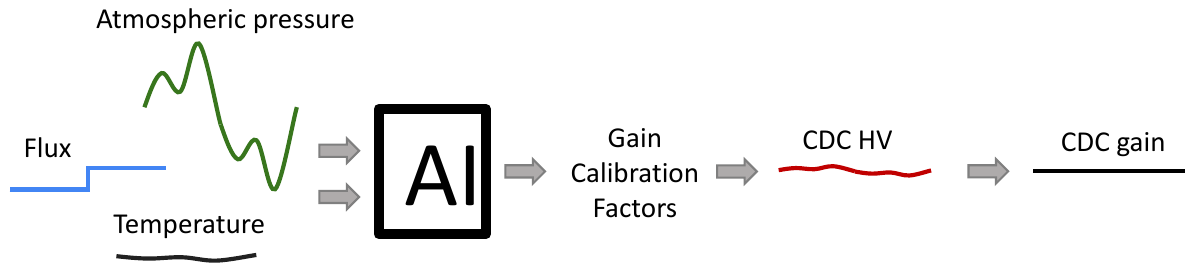}
    \caption{Diagram illustrating how environmental parameters are used to predict gain calibrations. The predicted calibrations are then used to determine adjustments to the detector high voltage to counterbalance these resulting in actual gain calibrations that are nearly constant over time.}
    \label{fig:AIEC_diagram}
\end{figure}

The GlueX Central Drift Chamber (CDC) consists of 3522 wires, each contained in a straw tube with a conductive coating\cite{ADHIKARI2021164807,JARVIS2020163727}. A high voltage (HV) of around 2125V is maintained between the wire in the center of the tube and the tube itself. This creates an intense electric field that accelerates electrons that have been liberated by the passing of charged particles towards the wire\cite{Sauli}. These electrons create a signal on the wire whose amplitude is related to the energy lost to the gas in the chamber by the initial charged particle moving through it. This amplitude can be used to help identify the exact type of that charged particle (see figure \ref{fig:CDC_PID}). The amplification of the signal or ``gain'' is determined by multiple factors which include the HV, atmospheric pressure, rate at which charged particles are passing through (due to beam flux $\times$ target thickness), and temperature of the gas. Traditionally, the data is analyzed after the experiment to determine the gain and calibration constants derived which are then used to correct the data during analysis. The goal of this project was to use a Machine Learning (ML) model to predict the calibration \textit{before} the data was taken using 3 of the parameters as inputs and then adjust the HV to counterbalance the effect on the gain. Figure \ref{fig:AIEC_diagram} illustrates this. The result would be to operate the detector in a more stable mode and significantly reduce the time needed for calibration after the data was taken.

\section{Initial System Development and Testing}

\begin{figure}[htb]
    \centering
    \includegraphics[width=1.0\linewidth]{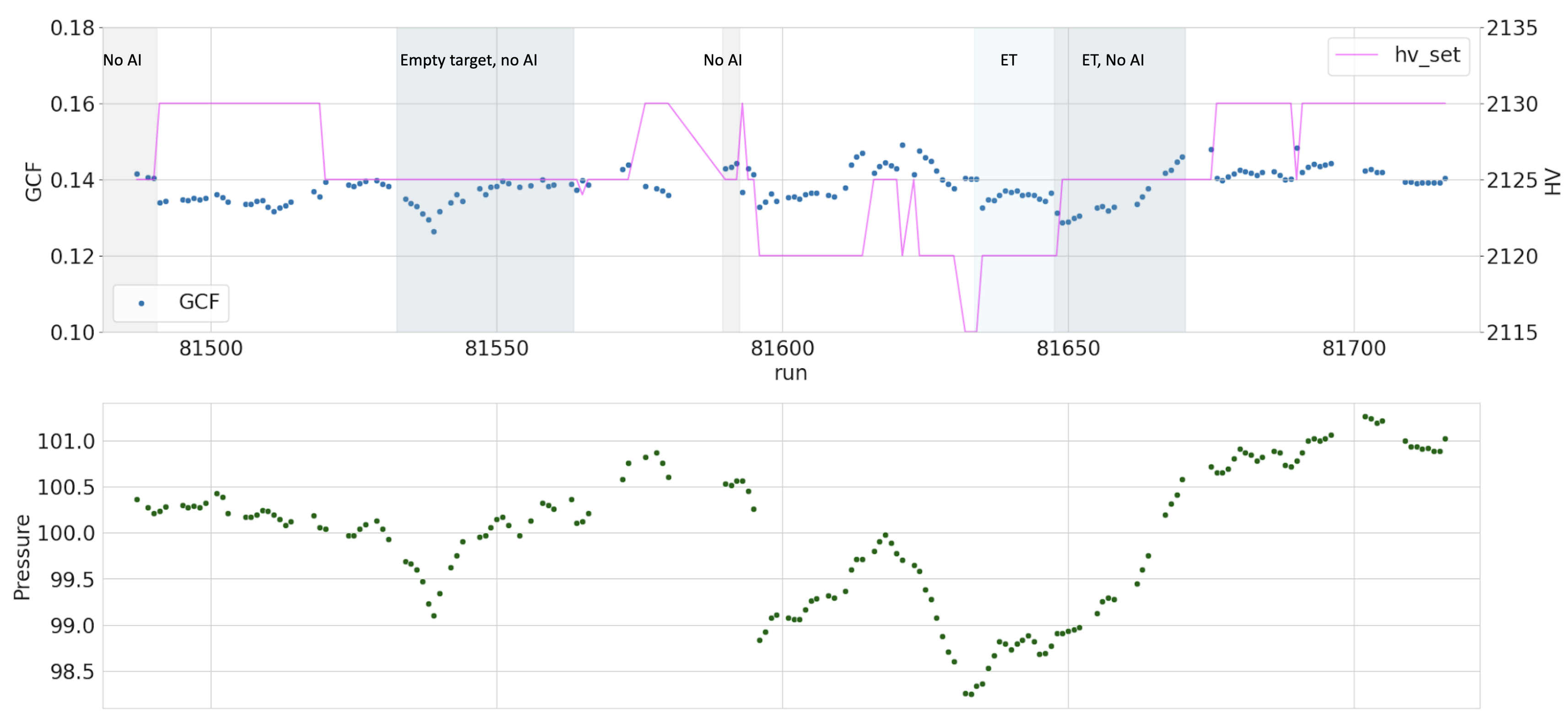}
    \caption{Plots from an initial semi-manual test with beam. This occurred during the first run of the PrimEx experiment in Fall of 2021. The x-axis of both plots is the run number which roughly correlates with time. The top plot shows the gain (blue dots plotted against left y-axis) and high voltage setting (magenta line plotted against right y-axis. The bottom plot shows the atmospheric pressure during this same time period. }
    \label{fig:AIEC_semi_manual}
\end{figure}

Development on the system began in 2021. A model was trained using historic GCF calibrations (derived from recorded data\cite{JONES2006366}) and environmental parameters as read from just before the data was acquired\cite{mcspadden_MLPS}. A Python script was used to gather current environmental conditions, execute the model, and convert the results into a suggestion for a new HV setting. This was done manually and it was up to shift takers to actually set the HV using the standard controls GUI. Forcing a human in the pipeline was done as a precaution and a way to assure collaborators the system would not risk data quality. Figure \ref{fig:AIEC_semi_manual} shows the HV and estimated gain correction factors (GCF) as a function of run number. The atmospheric pressure is also shown to help illustrate its correlation to the gain. For this test HV values were only changed by increments of 5V. This initial test was successful at improving the stability of the GCF.

\begin{figure}[htb]
    \centering
    \includegraphics[width=0.7\linewidth]{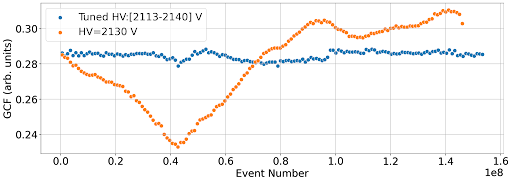}
    \caption{(Reproduced from \cite{mcspadden_MLPS}) Plot showing the first automated test of the AI/ML controlled system during a cosmic data run. The x-axis is event number which roughly corresponds to time over a 2 week period. The orange dots are from the half of the detector that was not controlled by the AI/ML model while the blue dots show the much more stable behavior of the half of the detector that was controller by the AI/ML model.}
    \label{fig:CDC_cosmics_plot}
\end{figure}

The next step involved automating the system so that it would not require a human. This was done during a period when beam was not available and instead used signals from cosmic rays. The policy was changed to allow 1V changes in the HV as opposed to the 5V policy used for the initial test. The automated system was allowed to modify the voltage on half of the wires automatically every 5 minutes over a 2 week period. The other half of the wires were maintained at constant voltage during this same time period. Guardrails were implemented to prevent the ML model from setting the HV outside of a limited range that was deemed safe under all circumstances. Figure \ref{fig:CDC_cosmics_plot} shows the GCF based on a later analysis of the data for both halves of the detector. The half controlled by the ML model had significantly more stable GCF values.

\section{Gaussian Process Regression}

The ML model used for the project was based on Gaussian Process Regression (GPR)\cite{NIPS1995_7cce53cf}. This was chosen as it naturally provides a value quantifying the uncertainty of its prediction.
A GPR is a non-parametric approach for regression that assumes the outputs and input variables follow a multivariate Gaussian distribution. For a single target GPR, the mean of this distribution is the predicted output, and the covariance captures the uncertainty associated with the prediction. 
An underlying function $R^d \rightarrow R$ maps training input, \textbf{$X$}, and corresponding targets $y$. 
The mapping uses a chosen covariance function, $k(\cdot,\cdot)$, which determines the smoothness of the predicted function between any two data points ($x$ and $x'$).
This function is used to construct the Gram matrix, $K$, which represents pairwise similarities between all data points in the training data:
\begin{align*} 
K_{n,n'} = k(x^n, x^{n'}), 
\end{align*}

where k, when applied to $X$ produces an $N$ x $N$ matrix, where the dataset is of size $N$. 

Our work employs a kernel function, $k(\cdot,\cdot)$ constructed as the \textbf{sum} of Scikit-learn's \cite{scikit-learn} Squared Exponential (RBF) and White Noise kernels.
The RBF kernel, Eq. \ref{eq:K_RBF_def}, captures the covariance and inherent smoothness between data points, while the White Noise kernel,  Eq. \ref{eq:K_WN}, accounts for the overall noise level present in the data.

\begin{equation} \label{eq:K_RBF_def}
{k_{RBF}}(x,x') = \exp\left(-\frac{(|x - x'|)^2}{2l^2}\right),
\end{equation}

where $l$ is the learned length-scale parameter used to scale the difference in distance between training observations.

\begin{equation}\label{eq:K_WN}
    k(x,x') = \sigma^2 I_n,
\end{equation}

where $\sigma^2$ is the variance of the noise and $I_n$ is the identity matrix.

\section{Integration with Controls System}

\begin{figure}[htb]
    \centering
    \includegraphics[width=0.95\linewidth]{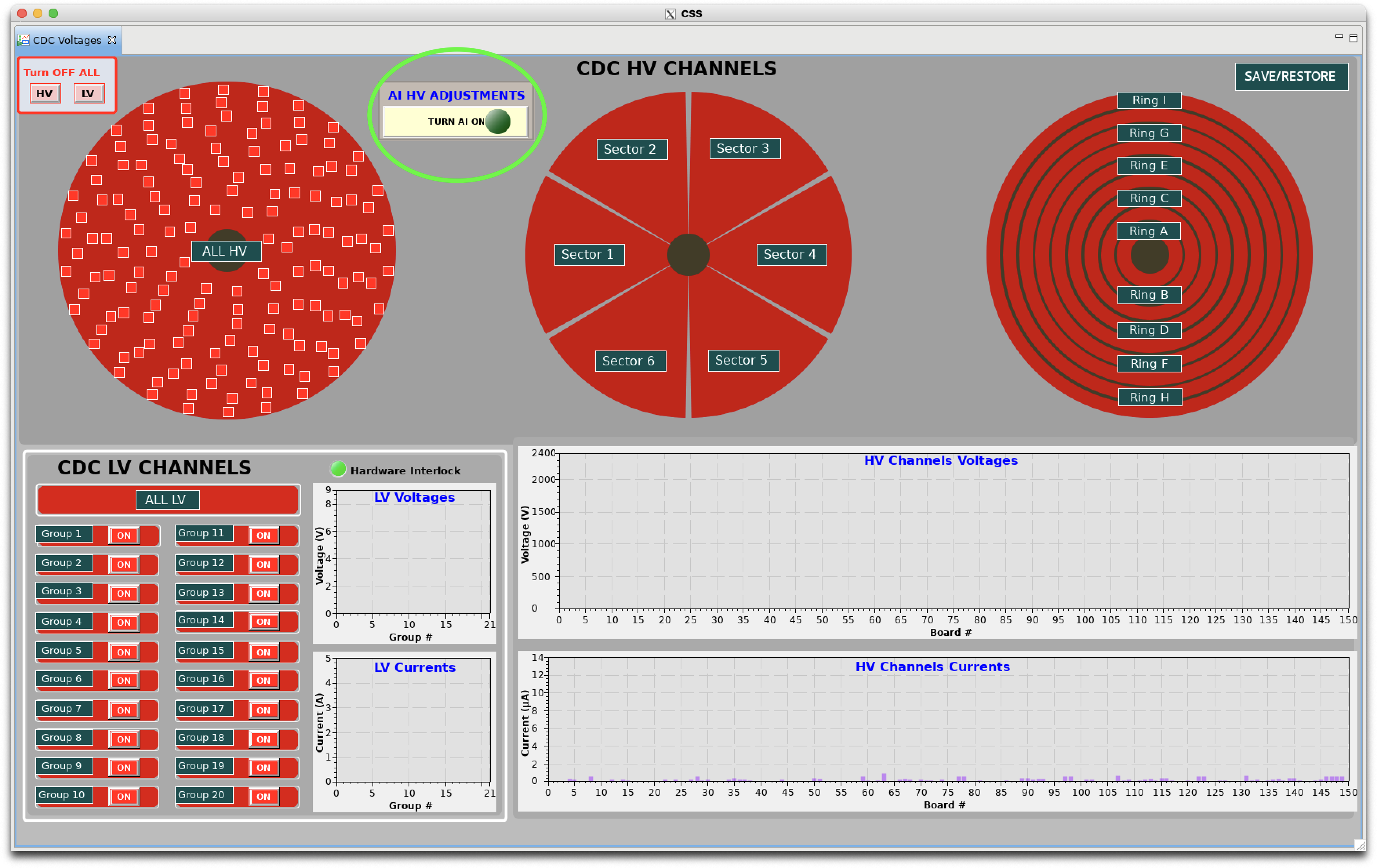}
    \caption{The CDC detector controls GUI. For this project, the button indicated by the green circle was added to allow the AI/ML control of the detector to be easily turned on/off at any time by the shift workers.}
    \label{fig:CDC_epic_GUI}
\end{figure}

Before deploying the system in a running experiment, a mechanism was integrated that would easily allow shift workers to turn ML control of the CDC on off. This included a new button on the standard control GUI for the CDC detector as shown in figure \ref{fig:CDC_epic_GUI}. Shift workers can range from graduate students to senior professors and from seasoned shift takers to novices. Data taking occurs 24/7 during a running experiment so system experts are not able to monitor this continuously. Placing an on/off switch in an easily accessible location and updating the shift worker documentation was considered necessary for such a new system. When the system was turned off, the ML model was still active and the recommended HV still recorded in a database. The HV itself was just not modified by the ML system.

\begin{figure}[htb]
    \centering
    \includegraphics[width=0.5\linewidth]{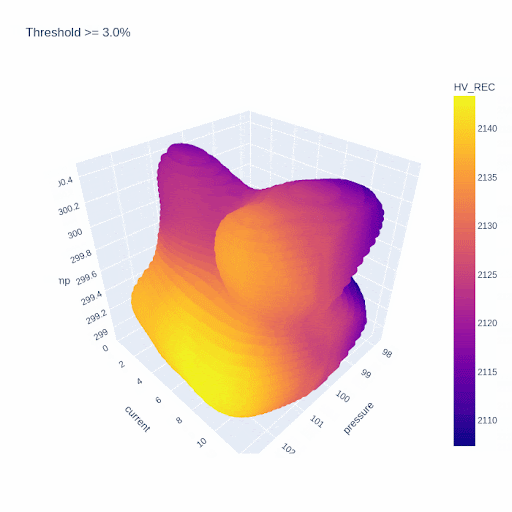}
    \caption{Visualization of the region of certainty for the AI/ML model. Points within the volume have an uncertainty of $\le3\%$. The AI/ML model is not allowed to control the detector high voltage for points outside of this surface. (See text for details.) }
    \label{fig:AIEC_error_pancakes}
\end{figure}

The ML model was trained on historic calibrations which included several regions of the input feature space. This did not cover all possible regions so the uncertainty quantification of the ML model output was needed to inform a policy that could make decisions on what to setting to use. Figure \ref{fig:AIEC_error_pancakes} shows a 3D rendering of the $3\%$ surface of model uncertainty. The final policy deployed will use the model recommendation for points within this $3\%$ surface but for points outside of it will revert to observation mode which automatically sets the HV to its default value of 2125V. Data gathered while in observation mode will contribute to future model training causing the surface to increase as more areas of the feature space are encountered.

\section{Results from Production Running}

\begin{figure}[htb]
    \centering
    \includegraphics[width=0.9\linewidth]{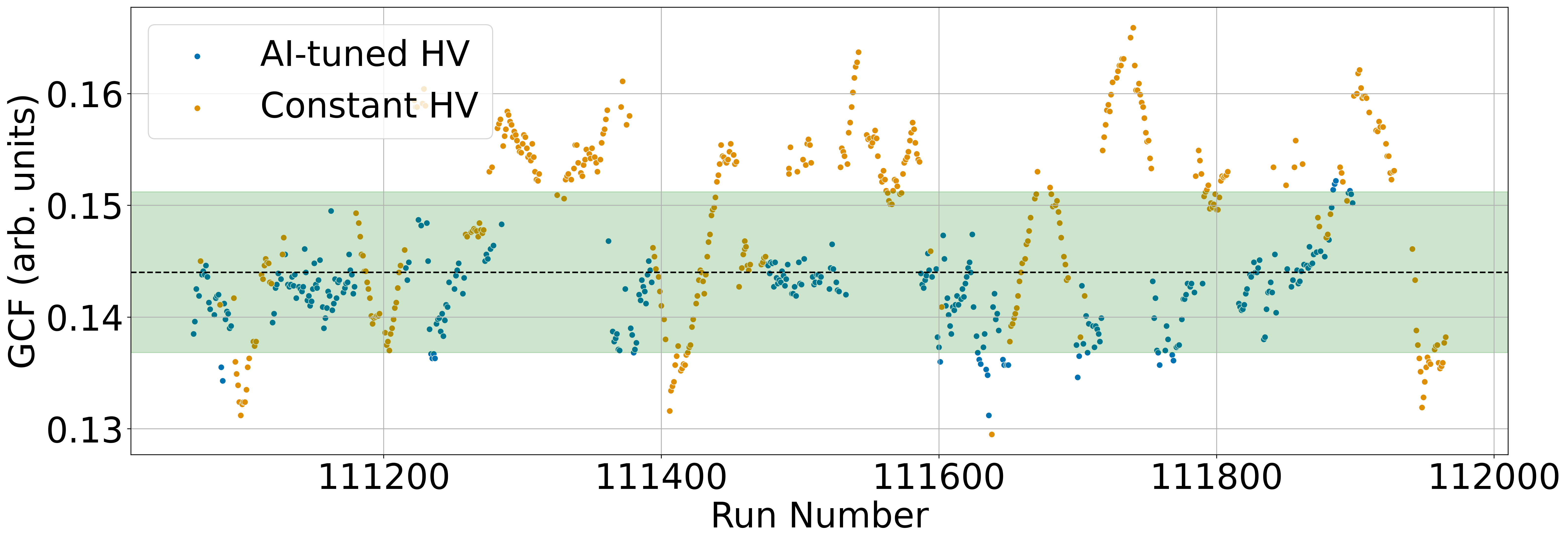}
    \caption{Gain correction factor (GCF) as a function of time (Run Number) for the PrimEx experiment's second run during the Fall of 2022. Orange points were taken at a constant 2125 V while blue points were taken with an AI/ML tuned HV setting. The dashed line indicates the ideal GCF while the green box corresponds to $\pm$ 5\% of that.}
    \label{fig:primex22_performance}
\end{figure}

The fully automated system is now deployed as part of standard production for the GlueX detector. Figure \ref{fig:primex22_performance} shows the GCF for the second run of the PrimEx experiment. This experiment included running conditions at the edge of the feature space on which the ML model was trained. Thus, it includes several regions where the ML system dropped to observation mode and used a constant HV setting. The green region indicates the $\pm5\%$ band that was an initial goal of system.

\begin{figure}[htb]
    \centering
    \includegraphics[width=0.9\linewidth]{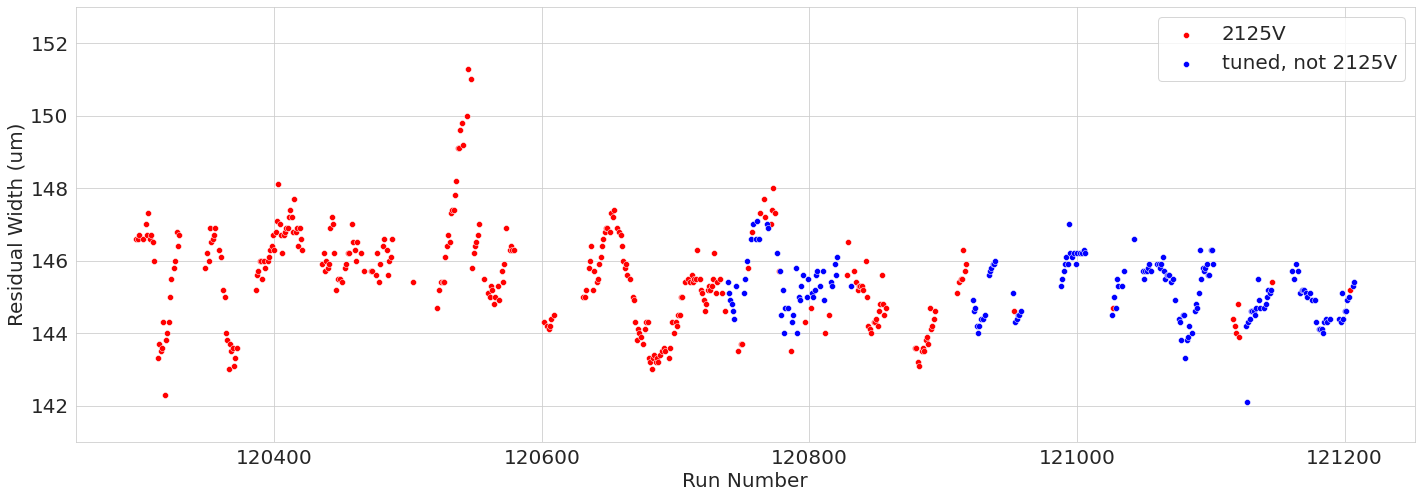}
    \includegraphics[width=0.9\linewidth]{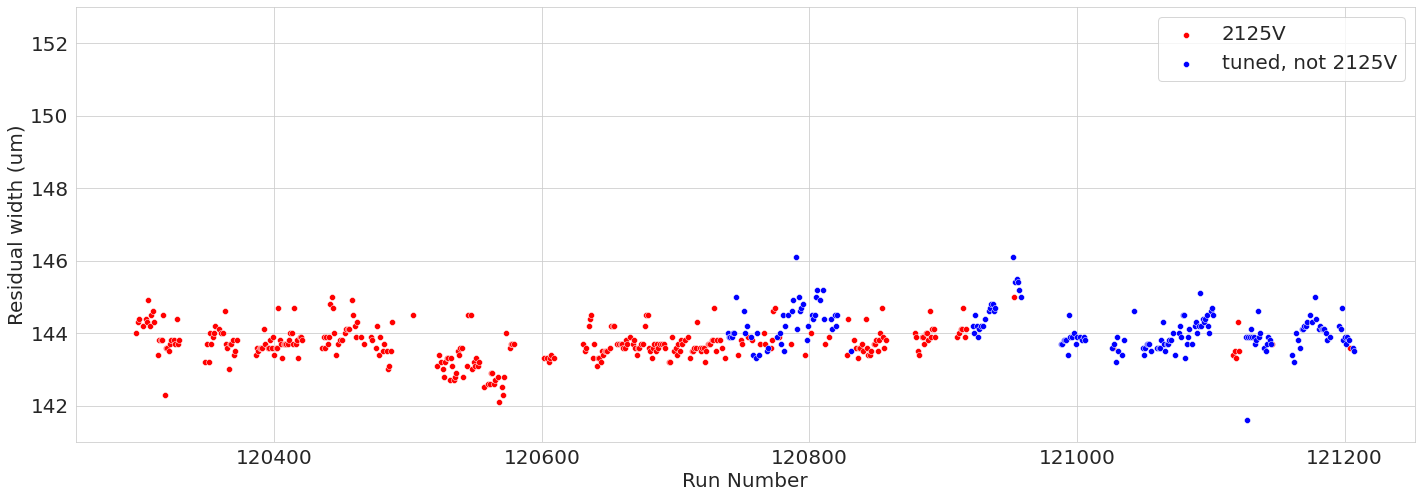}
    \caption{\textit{Data taken during GlueX experiment phase II in Spring of 2023}\\
    \textbf{TOP:} Widths of residual time-to-distance(TtoD) distributions obtained before calibration for runs taken at 2125 V (red) and a HV setting determined by the AI/ML system (blue). This shows that the adjustments made by the AI/ML model to stabilize the gain did not introduce instability to the TtoD calibration. \\
    \textbf{BOTTOM:} Widths of residual time-to-distance(TtoD) distributions obtained using a linear function dependent only on gas density. This new, fast technique for calibrating the TtoD was developed after noticing an interesting correlation while working on the gain calibrations.}
    \label{fig:ttod_beforeCalibration}
\end{figure}

A final concern with this system was that it was not clear if stabilizing the GCF calibration for the CDC would lead to less stability in the other calibration constants for the detector that are used to determine the time-to-distance(TtoD) conversion. Figure \ref{fig:ttod_beforeCalibration} shows the TtoD residual width as a function of run. The plots include periods of both constant HV (red points mostly to the left of the plot) and ML controlled HV (blue points mostly to the right of the plot). The top plot indicates that the ML controlled period was no less stable than the period using the legacy mode of running with constant HV. The bottom plot shows the residual widths after applying a correction based on the gas density. This correction was derived as a byproduct of this project which exposed a correlation in the TtoD that had not been noticed before. This further reduced the time needed to fully calibration the detector.

\section{Summary}

A system utilizing an ML model to automatically control the High Voltage of the GlueX Central Drift Chamber detector has been deployed in production experiments.
The system predicts calibrations based on environmental factors available prior to taking data. The predictions are then used to adjust the HV in order to stabilize the gain of the detector. The system was developed in stages to ensure safe, robust operation and to instill confidence and trust in the scientists whose data depended on it.

\acknowledgments

This work is supported by a grant from the U.S. Department of Energy, Office of Science, Office of Nuclear Physics under the LAB-20-2261 FOA.\\
\newline
\noindent
The Carnegie Mellon Group is supported by the U.S. Department of Energy, Office of Science, Office of Nuclear Physics, DOE Grant No. DE-FG02-87ER40315\\
\newline
\noindent
This research used resources of the Thomas Jefferson National Accelerator Facility, which is a DOE Office of Science User Facility supported by the U.S. Department of Energy, Office of Science, Office of Nuclear Physics under contract DE-AC05-06OR23177.






\bibliographystyle{plainnat}

\bibliography{biblio}




\end{document}